\documentclass[iop,apj,revtex4]{emulateapj}
\usepackage{graphicx}
\usepackage{amsmath,amssymb}

\begin{document}
\title{Solar wind protons at 1 AU: trends and bounds, constraints
and correlations}

\author{Petr Hellinger\altaffilmark{1}}
\email{petr.hellinger@asu.cas.cz}
\author{Pavel M. Tr\'avn\'\i\v cek\altaffilmark{2,1}}

\altaffiltext{1}{Astronomical Institute, AS CR,
Bocni II/1401,CZ-14100 Prague, Czech Republic
}
\altaffiltext{2}{Space Sciences Laboratory, UCB, Berkeley, USA}

\begin{abstract}
The proton temperature anisotropy in the solar wind exhibits apparent bounds which are 
compatible with the theoretical constraints imposed by temperature-anisotropy driven kinetic instabilities. 
Recent statistical analyses based on conditional averaging indicate that near these theoretical constraints
 the solar wind protons have typically enhanced temperatures and a weaker collisionality. 
Here we carefully analyze the solar wind data and show that these results are a 
consequence of superposition of multiple correlations in the solar wind, namely, 
they mostly result from the correlation between the proton temperature and the 
solar wind velocity and from the superimposed anti-correlation between 
the proton temperature anisotropy and the proton parallel beta in the fast solar wind. 
Colder and more collisional data are distributed around temperature isotropy  whereas
 hotter and less collisional data  have a wider range of the temperature anisotropy
anti-correlated with the proton parallel beta
with signatures of constraints owing to the temperature-anisotropy driven instabilities.
However, most of the hot and weakly collisional data, including the hottest and least collisional ones,
 lies far from the marginal stability regions.
Consequently, we conclude that there is no clear relation between the enhanced temperatures 
and instability constraints and that the conditional averaging used for these analyses
must be used carefully and need to be well tested. 
\end{abstract}
\pacs{?}
\maketitle

\section{Introduction}
Physical mechanisms responsible for acceleration and heating of the solar wind plasma
still remain to a large extent an open problem \citep{holl08,mave11,hellal13}. 
Different processes leave  imprints in the solar wind
particle properties and may be thus possibly identified \citep{mars12,mattal12}.
The solar wind protons at 1 AU exhibit many different correlations and apparent bounds
which are not fully understood.
The proton temperature $T_\mathrm{p}$ is correlated with the solar wind velocity $v_{sw}$
\citep{mattal06c,demo09,ellial12}
whereas the proton number density $n_\mathrm{p}$ is anti-correlated with $v_{sw}$.
The proton parallel beta $\beta_{\mathrm{p}\|}=2\mu_0 n_\mathrm{p} k_B T_{\mathrm{p}\|}/B_0^2$
(the ratio between the proton parallel pressure and the magnetic pressure) and the proton temperature anisotropy 
$\mathcal{A}_{\mathrm{p}}=T_{\mathrm{p}\perp}/T_{\mathrm{p}\|}$
are anti-correlated in the fast solar wind \citep{marsal04,hellal06}  as
\begin{equation}
\mathcal{A}_\mathrm{p} \sim 1.16
\beta_{\mathrm{p}\parallel }^{-0.55}.
\label{marsch}
\end{equation}
Moreover, $\beta_{\mathrm{p}\|}$ increases and $\mathcal{A}_{\mathrm{p}}$ decreases with the radial distance
between 0.3 and 1 AU 
following the trend of Eq.~(\ref{marsch})
 in the fast solar wind
 \citep{mattal07}. Here 
$T_{\mathrm{p}\|}$ and $T_{\mathrm{p}\perp}$ are
the proton parallel and perpendicular temperatures (with
respect to the ambient magnetic field), respectively,
$B_0$ is the magnitude of the ambient magnetic field,
$\mu_0$ and $k_B$ are the magnetic permeability and
the Botzmann constant, respectively.

Coulomb collisions are typically too weak to keep the plasma in thermal equilibrium and 
the solar wind protons exhibit important particle temperature anisotropies. 
The proton temperature anisotropy $\mathcal{A}_{\mathrm{p}}$, however,
seems to be constrained. The data distribution
in the $(\beta_{\mathrm{p}\|},\mathcal{A}_\mathrm{p})$ space
has roughly a rhomboidal shape and the apparent
bounds on the high $\beta_{\mathrm{p}\|}$ side  
are compatible with the
theoretical constraints imposed by kinetic instabilities
driven by the proton temperature anisotropy \citep{kaspal02,hellal06}.

Important solar wind parameters (such as the collisional time,
the proton temperature, and the amplitude of the turbulent/fluctuating magnetic field) seem to be
related to kinetic instabilities. Some statistical studies indicate that these
physical quantities are enhanced or reduced near marginal stability regions
of the temperature-anisotropy driven instabilities \citep{baleal09,marual11,osmaal12b,wickal13}.
This may indicate a connection between these kinetic instabilities and
processes which are likely responsible for the 
proton heating such as the magnetohydrodynamic turbulence \citep{mave11,osmaal13}.
However, these studies are based on a conditional averaging: the data are split
in bins in the 2-D space of $\beta_{\mathrm{p}\|}$ and $\mathcal{A}_{\mathrm{p}}$ with variable 
sizes and number of points and an averaging within the different bins  is used to get
a dependence of a third physical parameter on $\beta_{\mathrm{p}\|}$ and $\mathcal{A}_{\mathrm{p}}$.
This procedure is not trivial and its results need to be tested.
 In this letter we analyze in detail possible relations between the proton collisionality
and temperature, and $\beta_{\mathrm{p}\|}$ and $\mathcal{A}_{\mathrm{p}}$.
In particular we demonstrate that the reported enhancements of the proton temperature
near marginal stability regions (resulting from the bin-averaged procedure) are related to the structure of the data
distribution in the ($\beta_{\mathrm{p}\|}$, $\mathcal{A}_{\mathrm{p}}$,
$T_\mathrm{p}$) space.

\section{Data analysis}
Here we use a large statistical data set (about 4 millions data points) 
 from the WIND spacecraft from 1995 to 2012 \citep{marual12}.
Let us first investigate a possible relation between 
$\beta_{p\|}$ and  $\mathcal{A}_{\mathrm{p}}$, 
  and the proton collisionality
characterized by 
the collisional time $\tau=\nu_T t_e$ (a proxy for the collisional age)
where $t_e= 1\ \mathrm{AU}/v_{sw}$ is the expansion (transit) time
and $\nu_T$ is the collisional proton isotropization frequency
defined as 
\begin{equation}
\left[\frac{\mathrm{d}(T_{\mathrm{p}\perp}-T_{\mathrm{p}\|})}{\mathrm{d}t}\right]_{collisions}=-\nu_T(T_{\mathrm{p}\perp}-T_{\mathrm{p}\|});
\end{equation}
$\nu_T$ may be derived assuming a bi-Maxwellian velocity distribution function and expressed in terms of the standard Gauss hypergeometric function $\ _2F_1$ as \citep{hetr09,hetr10}  
\begin{equation}
\nu_{T}=
\frac{ e^4 n_\mathrm{p} \ln\Lambda}
{30\pi^{3/2}\varepsilon_0^2 \sqrt{m_\mathrm{p} k_{B}^{3} T_{\mathrm{p}\|}^{3}} }
\ _2F_1 \left(2,\frac{3}{2}; \frac{7}{2};1-\mathcal{A}_\mathrm{p}\right),
\end{equation}
where $e$ is the proton charge, $\ln\Lambda$ is the Coulomb logarithm, $\varepsilon_0$
is the permittivity of vacuum, and $m_\mathrm{p}$ is the proton mass. 
These four parameters are not however independent, they depend on the proton (parallel) temperature;
if we neglect the role of other parameters we get
\begin{align}
\mathcal{A}_{\mathrm{p}}\propto 1/ \beta_{p\|}\ \ \textrm{and} \ \
\tau \propto T_{\mathrm{p}}^{-3/2}. 
\label{dependent}
\end{align}
These interdependencies need to be taken in account when interpreting the observations; note that
while Eq.~(\ref{dependent}) predicts an anti-correlation between $\mathcal{A}_{\mathrm{p}}$ and $\beta_{p\|}$
it is not sufficient to explain quantitatively the observed anti-correlation of Eq.~(\ref{marsch}).

\begin{figure}[htb]
\centerline{\includegraphics[width=8cm]{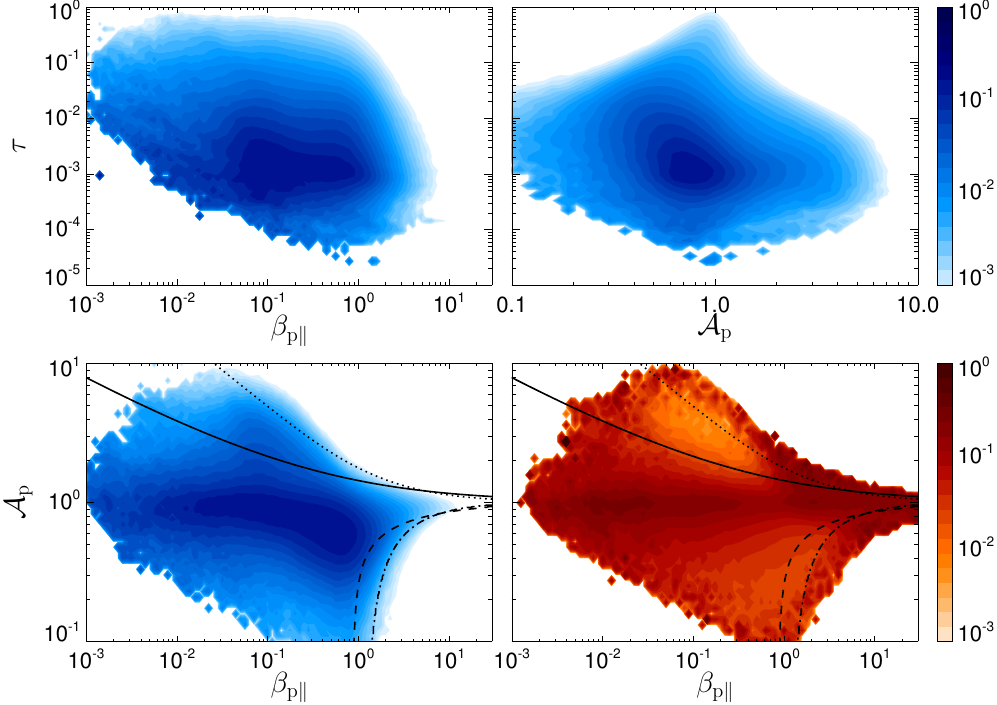}}
\caption{
The left top panel shows a color scale plot of the observed relative frequency 
of ($\beta_{\mathrm{p}\|}$, $\tau$), the right top panel shows a similar plot for
($\mathcal{A}_\mathrm{p}$, $\tau$), and the left bottom panel displays the case of ($\beta_{\mathrm{p}\|}$,$\mathcal{A}_\mathrm{p}$).
The right bottom panel shows the bin-averaged 
 $\tau=\tau(\beta_{\mathrm{p}\|},\mathcal{A}_\mathrm{p})$.
The overplotted curves on the bottom panels show the marginal stability relations
($\gamma_{max} = 10^{-3} \omega_{c\mathrm{p}}$) for the  (solid) proton cyclotron
(dotted) mirror, (dashed) parallel fire hose, and (dash-dotted) oblique fire hose
instabilities. 
\label{bpannut}}
\end{figure}

A natural way to investigate a relation between three physical quantities
would be a three-dimensional frequency plot/histogram. Such three dimensional
plots are however hard to visualize and interpret so that we use here only two-dimensional
analyses. In this way, the connections between the parameters due to interdependences
such as in Eq.~(\ref{dependent}) may be discern.
The relation between the proton  $\beta_{\mathrm{p}\|}$, the proton temperature anisotropy $\mathcal{A}_\mathrm{p}$
and the collisional time $\tau$ is investigated in Figure~\ref{bpannut}.
The left top panel shows a color scale plot of the observed relative frequency (normalized to
the bin size \citep{marual12}) of ($\beta_{\mathrm{p}\|}$, $\tau$), the right top panel shows a similar plot for
($\mathcal{A}_\mathrm{p}$, $\tau$), and the left bottom panel displays the case of ($\beta_{\mathrm{p}\|}$,$\mathcal{A}_\mathrm{p}$).

The right bottom panel shows the bin-averaged collisional time $\tau$
 as a function of $\beta_{\mathrm{p}\|}$ and $\mathcal{A}_\mathrm{p}$ \citep{baleal09}:
for a given bin in $\beta_{\mathrm{p}\|}$ and $\mathcal{A}_\mathrm{p}$ the average
collisional time $\tau$ is calculated and the result is plotted as $\tau=\tau(\beta_{\mathrm{p}\|},\mathcal{A}_\mathrm{p})$.
Note that different bins have different sizes and different ``weights'', i.e., different
number of data points used for averaging. Here we show results only for bins with more than
20 data points.

The bottom left panel shows the data distributed in the
 ($\beta_{\mathrm{p}\|}$,$\mathcal{A}_\mathrm{p}$) space which
is useful for the electromagnetic temperature-anisotropy driven instabilities;
the different overplotted curves (solid one for the proton cyclotron instability,
dotted one for the mirror instability, dashed one for the parallel fire hose,
and the dash-dotted one for the oblique fire hose)
 denote the marginal stability relations,
i.e., where the corresponding bi-Maxwellian linear kinetic theory predicts 
that the fastest growing mode has
the growth rate $\gamma_\mathrm{max}(\beta_{\mathrm{p}\|},\mathcal{A}_\mathrm{p})=10^{-3}\omega_{c\mathrm{p}}$ 
(where $\omega_{c\mathrm{p}}$ is the proton cyclotron frequency)
\citep{hellal06}. The system becomes more unstable when
increasing $\beta_{\mathrm{p}\|}$ and/or when increasing (decreasing)
$\mathcal{A}_\mathrm{p}$ for $\mathcal{A}_\mathrm{p}>1$ (for $\mathcal{A}_\mathrm{p}<1$).
The data exhibit bounds which
are more compatible with theoretical constraints imposed by the 
weaker, oblique instabilities (mirror and oblique fire hose) apparently
at odds with the expected important role of the linearly dominant 
instabilities (proton cyclotron and parallel fire hose);
however,
the theoretical prediction is based on simplified and idealized plasma composition and properties \citep{mattal12}
and, moreover, these instabilities are resonant, i.e., their
stability strongly depends on a detailed structure of the particle velocity
distribution function \citep{hetr11,isenal13}.

\begin{figure}[htb]
\centerline{\includegraphics[width=8cm]{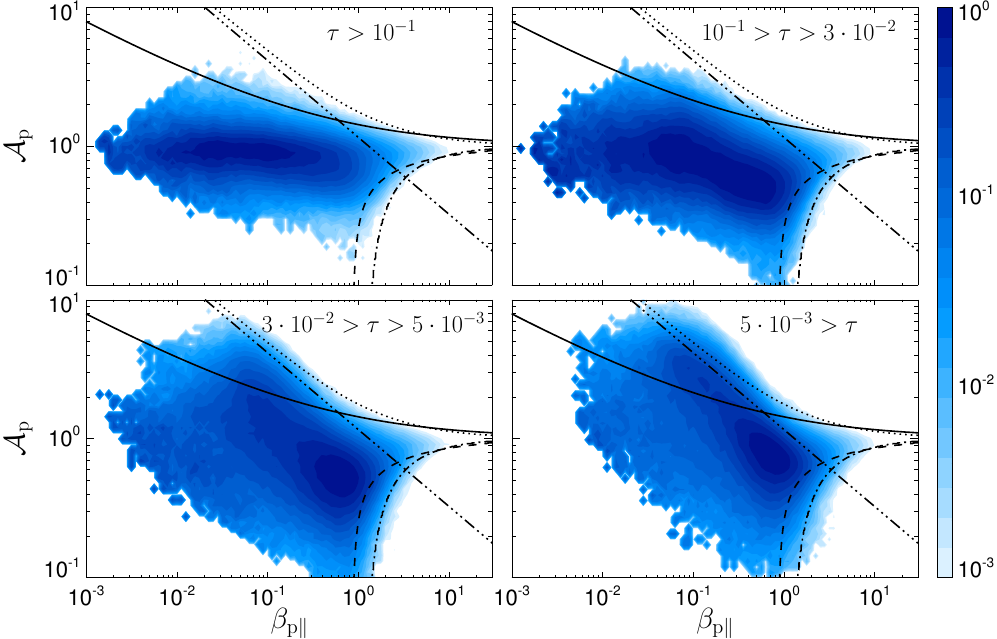}}
\caption{Color scale plots of the relative frequency of ($\beta_{\mathrm{p}\|}$,$\mathcal{A}_\mathrm{p}$)
for different ranges of collisional time (top left)
$\tau>10^{-1}$, (top right) 
$10^{-1}>\tau>3\cdot 10^{-2}$, (bottom left),
 $3\cdot 10^{-2}>\tau>5\cdot 10^{-3}$,
and (bottom right)
 $\tau< 5\cdot 10^{-3}$.
The overplotted curves show the marginal stability relations as in Fig.~\ref{bpannut}
whereas the dash-dot-dot-doted line displays the anti-correlation (\ref{marsch}).
\label{bpannut2}}
\end{figure}

Figure~\ref{bpannut} demonstrates that there is no clear relation between 
the collisional time $\tau$ and $\beta_{\mathrm{p}\|}$ (top left) whereas
for larger $\tau$ the proton temperature anisotropy $\mathcal{A}_\mathrm{p}$ is weaker and
approaches $\mathcal{A}_\mathrm{p}\sim1$ for very large $\tau$ (top right). 
The bottom right panel indicates that the data around temperature
isotropy  $\mathcal{A}_\mathrm{p}\sim1$  are on average more collisional whereas
the data near the marginal stability regions are less collisional \citep{baleal09}.

To understand this behavior we now analyze the relation between
$\tau$, $\beta_{\mathrm{p}\|}$, and $\mathcal{A}_\mathrm{p}$ in more detail. 
Figure~\ref{bpannut2} shows 
color scale plots of the relative frequency of ($\beta_{\mathrm{p}\|}$, $\mathcal{A}_\mathrm{p}$)
for different ranges of collisional times from more to less collisional plasmas
(from left to right and from top to bottom). Each range of $\tau$ has about
the same number of points (1 million). 
The dash-dot-dot-doted line displays Eq.~(\ref{marsch}).
We observe that
the most collisional protons exhibit a distribution centered around temperature isotropy
$\mathcal{A}_\mathrm{p}\sim 1$ with a large variation of $\beta_{\mathrm{p}\|}$.
This distribution gradually transforms  to an  anti-correlation between
$\beta_{\mathrm{p}\|}$ and $\mathcal{A}_\mathrm{p}$ with varying slopes.
For the lowest collisional times the
anti-correlation becomes comparable to Eq.~(\ref{marsch}) \citep{hellal06,mattal07}.
Figure~\ref{bpannut2} gives a clear explanation of the bottom right panel
of Fig.~\ref{bpannut}. The more collisional data 
contribute to the bin averages more around $\mathcal{A}_\mathrm{p}\sim 1$ (for lower  $\beta_{\mathrm{p}\|}$)
whereas  the less collisional data contribute more around
the anti-correlation (and further away from the isotropic region). Figure~\ref{bpannut}
then indicates lower average collisional time near marginal stability regions
as a result of the bin averaging, but the more detailed analysis of Fig.~\ref{bpannut2}
does not show  any clear relation between the reduced collisional age and marginal stabilities;
weakly collisional data are bounded by the marginal stability regions but most
of them lie far from them in vicinity of temperature isotropy.

It is also noteworthy that the data distribution in Fig.~\ref{bpannut2}
extends to lower $\beta_{\mathrm{p}\|}$ (and more isotropic $\mathcal{A}_\mathrm{p}$)
 for more collisional plasmas. The apparent bounds on the left hand side
(Fig.~\ref{bpannut}, bottom left, for low $\beta_{\mathrm{p}\|}$) 
of the data distribution in ($\beta_{\mathrm{p}\|}$, $\mathcal{A}_\mathrm{p}$)
are therefore likely a consequence of (proton-proton) Coulomb collisions.

\begin{figure}[htb]
\centerline{\includegraphics[width=8cm]{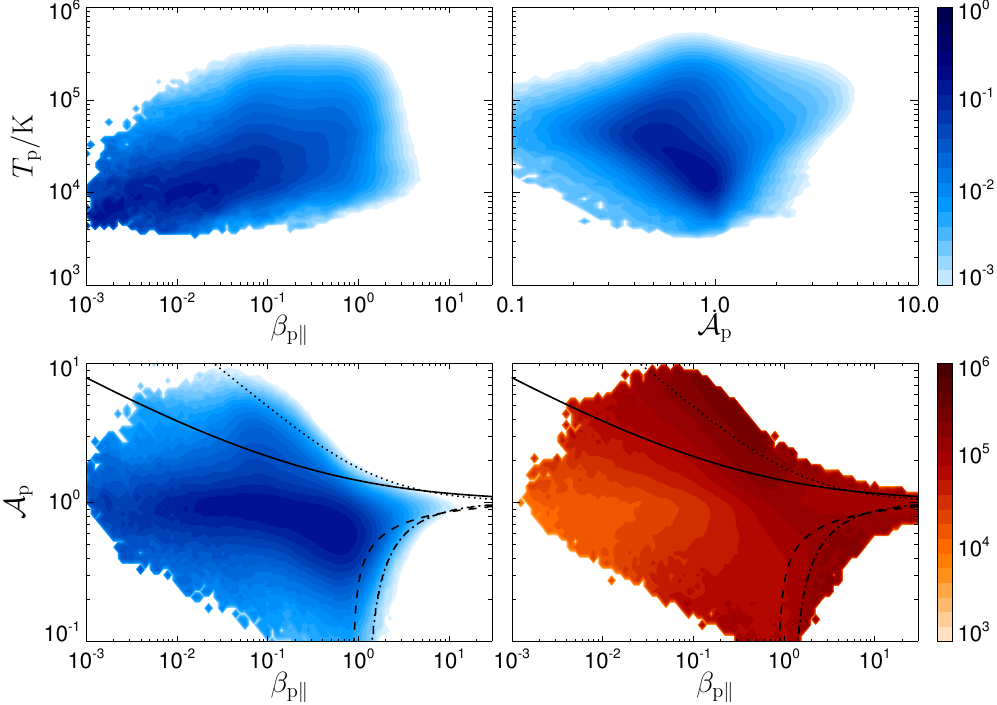}}
\caption{The left top panel shows a color scale plot of the observed relative frequency
of ($\beta_{\mathrm{p}\|}$, $T_\mathrm{p}$), the right top panel shows a similar plot for
($\mathcal{A}_\mathrm{p}$, $T_\mathrm{p}$), and the left bottom panel displays the case of ($\beta_{\mathrm{p}\|}$, $\mathcal{A}_\mathrm{p}$).
The right bottom panel shows the bin-averaged 
 $T_\mathrm{p}=T_\mathrm{p}(\beta_{\mathrm{p}\|},\mathcal{A}_\mathrm{p})$.
The overplotted curves on the bottom panels show the marginal stability relations
as in Fig.~\ref{bpannut}.
\label{bpant}}
\end{figure}

The proton temperature $T_\mathrm{p}$ exhibits  similar (but opposite) behavior with respect
to the  marginal stability regions in the $(\beta_{\mathrm{p}\|},\mathcal{A}_\mathrm{p})$
space compared to $\tau$ as expected from Eq.~(\ref{dependent}.
 $T_\mathrm{p}$ seems to be enhanced near the marginal stability regions \citep{marual11}. 
Let us now apply a similar analysis to the relation
between $\beta_{\mathrm{p}\|}$, $\mathcal{A}_\mathrm{p}$ and $T_\mathrm{p}$.  
Figure~\ref{bpant} shows
 color scale plots of the observed relative frequency
of ($\beta_{\mathrm{p}\|}$, $T_\mathrm{p}$) (top left), 
of ($\mathcal{A}_\mathrm{p}$, $T_\mathrm{p}$) (top right), and of ($\beta_{\mathrm{p}\|}$,$\mathcal{A}_\mathrm{p}$)
(bottom left).
The right bottom panel shows the bin-averaged proton temperature
 $T_\mathrm{p}$ as a function of $\beta_{\mathrm{p}\|}$ and $\mathcal{A}_\mathrm{p}$.
(only bins with more than
20 data points are shown).

\begin{figure}[htb]
\centerline{\includegraphics[width=8cm]{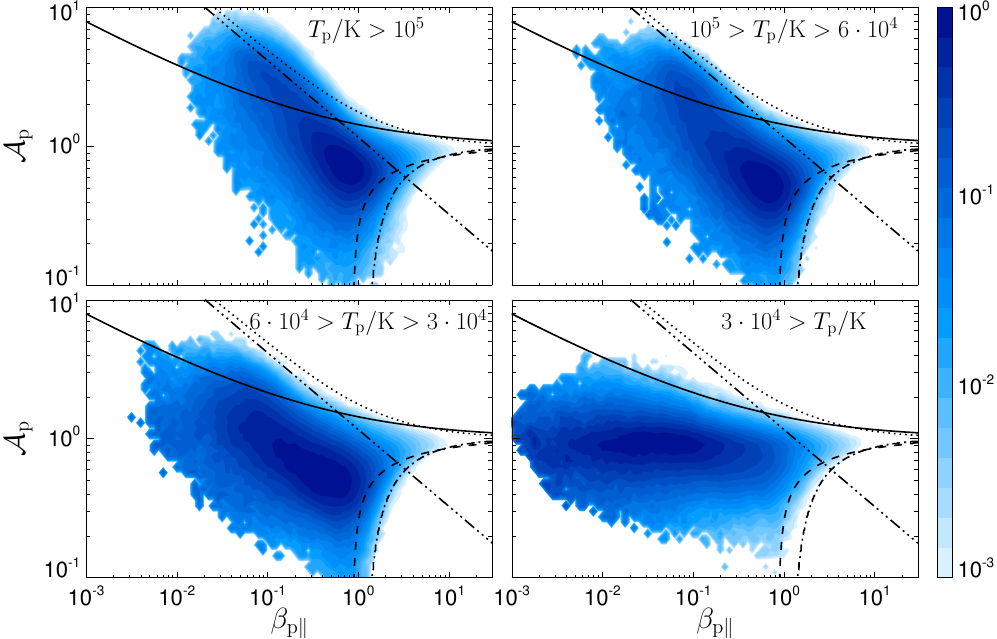}}
\caption{Color scale plots of the relative frequency of ($\beta_{\mathrm{p}\|}$, $\mathcal{A}_\mathrm{p}$)
for (top left) $T_\mathrm{p}/\mathrm{K}>10^{5}$,
(top right) $10^{5}>T_\mathrm{p}/\mathrm{K}>6\cdot 10^{4}$,
(bottom left)  $6\cdot 10^{4}>T_\mathrm{p}/\mathrm{K}>3\cdot 10^{4}$,
and (bottom right) $3\cdot 10^{4}>T_\mathrm{p}/\mathrm{K}$.
The overplotted curves show the marginal stability relations as in Fig.~\ref{bpannut}
whereas the dash-dot-dot-doted line displays the anti-correlation (\ref{marsch}).
\label{bpant2}}
\end{figure}

The proton temperature $T_\mathrm{p}$ indeed seems to be enhanced  near the
marginal stability regions. Let us now look in more detail at
the relation between the proton temperature and the  $(\beta_{\mathrm{p}\|},\mathcal{A}_\mathrm{p})$
space. Figure~\ref{bpant2} shows 
color scale plots of the relative frequency of ($\beta_{\mathrm{p}\|}$, $\mathcal{A}_\mathrm{p}$)
for different ranges of the proton temperature from hotter to colder protons
(from left to right and from top to bottom).
Each range of $T_\mathrm{p}$ has about
the same number of points (1 million).
 These partial data distributions
have clearly opposite behavior compared to that of the collisional time (Fig.~\ref{bpannut2}).
For the hottest protons the data distribution exhibit an anti-correlation 
similar to Eq.~(\ref{marsch}) which gradually transforms (anti-correlations with a decreasing slope) to
a distribution centered around temperature isotropy with a wide range
of $\beta_{\mathrm{p}\|}$ for the coldest protons.
Again, Figure~\ref{bpant2} gives a clear explanation of the bottom right panel
of Fig.~\ref{bpant}. Colder, more collisional protons contribute
importantly around temperature isotropy whereas hotter, less collisional protons contribute
more around the anti-correlation between $\beta_{\mathrm{p}\|}$ and $\mathcal{A}_\mathrm{p}$. 
As a result of the bin-averaging procedure the proton temperature seems to be
enhanced near marginal stability regions (and for important temperature anisotropies).
However, no clear relation between the enhanced proton temperature and marginal stabilities is found;
hotter proton data are bounded by the marginal stability regions but most
of them lie far from them in vicinity of temperature isotropy. 

The presented analysis divided the data in subsets (with about the same sizes) according to the proton collisional
age or temperature. This approach misses  smaller scale structure of the data. In order to
test whether we don't loose important properties we repeated the analysis by splitting the
data in $\tau$ and  $T_\mathrm{p}$ in 8 and also 16 subsets with about the same sizes
and we recovered essentially the same behavior, the transition from colder, collisional
data distributed around temperature isotropy  to hot, less collisional data exhibiting
an anti-correlation between $\beta_{\mathrm{p}\|}$ and $\mathcal{A}_\mathrm{p}$. In each case
even the hottest (and the least collisional) proton data are bounded by the marginal stability regions but most
of them lie far from them in vicinity of temperature isotropy.

\begin{figure}[htb]
\centerline{\includegraphics[width=8cm]{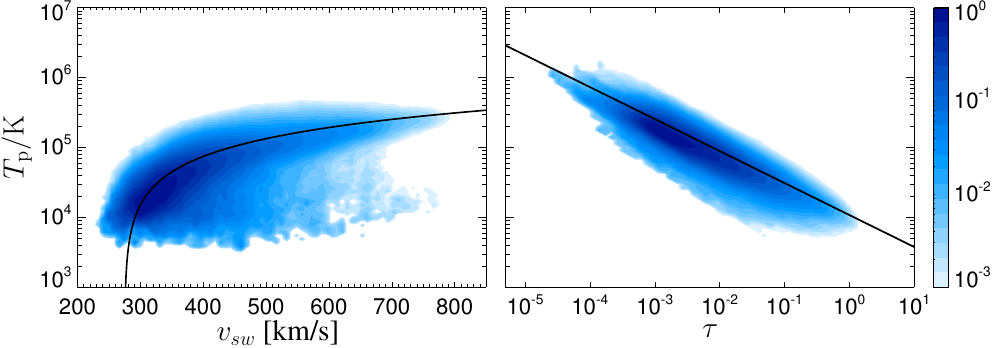}}
\caption{Color scale plots of the relative frequency of ($v_{sw}$, $T_\mathrm{p}$)
(left panel) and of ($\tau$, $T_\mathrm{p}$) (right panel).
Overplotted curves show Eq.~(\ref{ellial12}) and Eq.~(\ref{anticor}).
\label{nutt}}
\end{figure}

The proton temperature and the collisional time have clear opposite behaviors indicating
an anti-correlation between them. Such an anti-correlation is expected from Eq.~(\ref{dependent})
and, moreover, the proton temperature
correlates with the solar wind velocity $v_{sw}$ which in turn anti-correlates with the proton
density. Figure~\ref{nutt} confirms this property, it shows on the left panel
the relative frequency of ($v_{sw}$, $T_\mathrm{p}$) and that of ($\tau$, $T_\mathrm{p}$) on the right panel. 
The relation between $v_{sw}$ and  $T_\mathrm{p}$ is roughly linear 
\begin{equation}
T_\mathrm{p}/\mathrm{K} = -1.6 \cdot 10^5  +   590 \times v_{sw} / (\mathrm{km/s})
\label{ellial12}
\end{equation}
(shown by the solid line on Fig.~\ref{nutt}, left) and it is
similar to the trend observed by \citep{ellial12}.
The relation between the collisional time $\tau$ and $T_\mathrm{p}$ is
well described by
\begin{equation}
T_\mathrm{p}/\mathrm{K} = 1.1 \cdot 10^4    \tau^{-0.46}
\label{anticor}
\end{equation}
(shown by the solid line on Fig.~\ref{nutt}, right). 
The anti-correlation of Eq.~(\ref{anticor})
cannot be explained only by 
Eq.~(\ref{dependent});
taking $T_\mathrm{p}\propto  v_{sw}$ (roughly Eq.~(\ref{ellial12})) one gets 
$T_\mathrm{p} \propto   \tau^{-2/5}$ which is close to the observed Eq.~(\ref{anticor}).

\section{Discussion}
The solar wind protons exhibit many not-yet-fully-understood 
correlations. 
The proton temperature $T_\mathrm{p}$ is correlated with the solar wind velocity $v_{sw}$,
 the collisional time $\tau$ is well anti-correlated with $T_\mathrm{p}$, and
the proton parallel beta $\beta_{\mathrm{p}\|}$ and the proton temperature anisotropy
$\mathcal{A}_{\mathrm{p}}$
are anti-correlated in the fast solar wind.

The bin-averaging procedure produces plots
as $\tau=\tau(\beta_{\mathrm{p}\|}, \mathcal{A}_\mathrm{p})$
and $T_\mathrm{p}=T_\mathrm{p}(\beta_{\mathrm{p}\|}, \mathcal{A}_\mathrm{p})$
(see Figs.~\ref{bpannut2} and \ref{bpant2}, bottom right)
which indicate a connection between the marginal stability regions of kinetic
instabilities and the collisional time or the proton temperature \citep{baleal09,marual11}.
One of the problems of the bin averaging is that it combines averages 
over highly variable bin and data sizes. This effect may in some cases
help to discern some trends. However, here we show that reduced
$\tau$ and enhanced $T_\mathrm{p}$ near the marginal stability regions
 rather reflects the data structure in the corresponding three-dimensional
space connected with the multiple correlations in the solar wind;
there is no statistically significant number of data points with
reduced $\tau$ and enhanced $T_\mathrm{p}$ near the marginal stability regions.
Colder and more collisional data are distributed around temperature isotropy  whereas
 hotter and less collisional data  have a wider spread in the space $(\beta_{\mathrm{p}\|},\mathcal{A}_\mathrm{p})$
reaching the  marginal stability regions; 
however, most of the hot and weakly collisional data
(including the hottest and least collisional ones)
 lie far from the marginal stability regions.
We conclude that the bin-averaging method is a nontrivial procedure which should
be carefully used and its results
must be tested in detail which is not usually done. Some previous studies
where this method has been applied 
\citep{kaspal08,baleal09,osmaal12b,kaspal13,wickal13,osmaal13,boural13,serval14}
likely need to be revisited. In particular,
the fluctuating magnetic field $\delta B$ (3 second r.m.s.) 
dependence on $\beta_{\mathrm{p}\|}$ and $\mathcal{A}_\mathrm{p}$
(obtained through the bin-averaging method)
exhibits enhancements of $\delta B$ near the marginal stability regions
similarly to the proton temperature \citep{baleal09}.
However, the 3 second r.m.s. $\delta B$ is clearly anti-correlated with
the collisional time \cite[see Fig.~3 of][]{baleal09} so that we expect that the enhancements
of  $\delta B$   are to some extent a consequence of this anti-correlation.
On the other hand, \cite{wickal13} used Ulysses data
to show (using the bin-averaging method) that the fluctuating magnetic field
on ion scales is enhanced near the marginal stability regions. This analysis
used only data from the fast solar wind and therefore the influence of the correlation between
the solar wind velocity and the level of magnetic fluctuations is likely
negligible in this case.
Further studies are needed to understand origins and consequences of multiple correlations in
the solar wind.
 
\acknowledgments
Authors acknowledge the grant P209/12/2023 of the Grant Agency of the Czech Republic
and the project RVO:67985815.
The research leading to these results has received funding from the
European Commission's 7th Framework Programme under
 the grant agreement \#284515 (project-shock.eu).
Wind data were obtained from the NSSDC website
http://nssdc.gsfc.nasa.gov.
Authors also acknowledge  
discussions with S.~D.~Bale, J.~C.~Kasper, S. Landi, B.~A.~Maruca,
W.~H.~Matthaeus, and L. Matteini and thank the referee for useful comments.

\end{document}